\documentclass[12pt]{article}
\usepackage{graphicx}
\usepackage{amssymb}
\usepackage{amsfonts}
\usepackage{amsmath}
\usepackage{lscape}
\usepackage{epsfig}
\usepackage[english]{babel}

\begin{document}


\begin{center}
{\Large \bf Two-particle multiplets splitting as a guideline in nucleon pairing estimations} \\

\vspace{4mm}

{ L.T.\,Imasheva$^{a} $, B.S.\,Ishkhanov$^{a, b} $,  S.V.\,Sidorov$^{a} $, M.E.\,Stepanov$^{a} $,}\\    T.Yu.\,Tretyakova$^{b,}$\footnote{E-mail: tretyakova@sinp.msu.ru}\\

{\small $^{a}$\,Department of Physics, Lomonosov Moscow State University,  Moscow, Russia}

{\small $^{b}$\,Skobeltsyn Institute of Nuclear Physics, Lomonosov Moscow State University,  Moscow, Russia}
\end{center}
\begin{abstract}

The ground state multiplet structure for nuclei over the wide range of mass number $A$ was calculated in $\delta$-approximation and different mass relations for pairing energy was analysed in this work. Correlation between the calculated multiplet structure and experimental data offer a guideline in deciding between mass relations for nucleon pairing.
 
\end{abstract}
\vspace*{6pt}


\label{sec:intro}
\section*{Introduction}

The nucleon pairing in atomic nuclei leads to a systematic variation depending on the evenness or oddness of $Z$ and $N$. 
The even-odd  staggering (EOS) is an estimate of the pairing energy of identical nucleons. 
It can be determined from empirical masses of four \cite{BM} or five \cite{moller} adjacent isotopes:
\begin{equation}\label{four_point}
\Delta_n^{(4)}(N)=\frac{(-1)^N}{4}[-S_n(N+1)+2S_n(N)-S_n(N-1)],\\
\end{equation}
\begin{equation}\label{five_point}
\begin{array}{rl}
\Delta_n^{(5)}(N)&=1/2[\Delta_n^{(4)}(N)+\Delta_n^{(4)}(N+1)]=\\
&=(-1)^N/8[-S_n(N+2)+3S_n(N+1)-3S_n(N)+S_n(N+2)],
\end{array}
\end{equation}
where $S_n(N) = B(N) - B(N-1)$  is neutron separation energy from the nucleus $(N,Z)$, and $B(N)$ is the total nuclear binding energy. 
In (\ref{four_point}) and (\ref{five_point}) for neutron EOS the proton number $Z$ is fixed. 
Similar expressions (here and after) for protons can be obtained by fixing the neutron number $N$ and replacement in the expressions $N$ on $Z$.

Much research is devoted to the evaluation the pairing and mean-field contributions to the experimental EOS. In \cite{Dob} the EOS estimation based on the binding energies of three adjacent nuclei was suggested:
\begin{equation}\label{three_point}
\Delta_n^{(3)}(N)=\frac{(-1)^N}{2}[S_n(N)-S_n(N+1)],
\end{equation}
and it was shown that the expression (\ref{three_point}) for an odd neutron number $\Delta_n^{(3)}(N+1)$ is the best approach to the nucleon pairing. This conclusion is consistent with a direct definition of the pairing energy of two neutrons $\Delta_{nn}$ as the difference between the separation energy of the neutron pair $S_{nn}$ from the nucleus $(Z,N)$ and doubled separation energy of neutron $S_n$ from nucleus $(Z,N-1)$ \cite{Preston}:
\begin{equation}\label{nn}
\Delta_{nn}(N)=S_{nn}(N) - 2S_n(N-1) =S_n(N) - S_n(N-1) = 2\Delta_n(N-1) 
\end{equation}
The main objective of this study is to analyze the different EOS estimations with another nucleon pairing manifestation, namely, the formation of ground state multiplet (the distinctive set of levels in the spectrum of low-lying excited states with $J^P = 0^+, 2^+, \dots J_{\max}^+$).
\label{sec:model}
\section{The seniority model}

Analysis of EOS effect in different theoretical approaches is the subject of many studies. Following  \cite{Dob}, let us consider the expressions (\ref{four_point}-\ref{nn}) in the seniority model which describes the motion of $N$ nucleons in the $2\Omega$-fold degenerate shell. The energy eigenvalues in this model can be written in terms of the particle number $N$ and seniority $v$ -- the number of unpaired nucleons:
 \begin{equation} \label{E} 
E(N,v)=-\frac{1}{4}G(N-v)(2\Omega-v-N+2)
\end{equation}
where $G$ is the pairing parameter, $2\Omega = 2j+1$, for even nucleon number $N=2n$ seniority of ground state is $v=0$, for odd nucleon number $N=2n+1$ seniority of ground state is $v=1$.
In \cite{Dob} the expression for EOS effect defined by (\ref{three_point}) is obtained:
\begin{equation} \label{sen1}
\Delta_n^{(3)}(N)=
\begin{cases}
\frac12 G\Omega+\frac12 G & \text{for $N=2n$,} \\
\frac12 G\Omega & \text{for $N=2n+1$.}
\end{cases}
\end{equation}
Since this result does not depend on $N$, averagings (\ref{four_point}) and (\ref{five_point}) coincide in the seniority model:
\begin{equation} \label{sen3}
\Delta_n^{(5)}(N)= \Delta_n^{(4)}(N) = \frac{1}{2} G \Omega + \frac{1}{4} G, \quad \text{for $N=2n$ and $N=2n+1$.}
\end{equation}
Expressions for direct determination of pairing energy $\Delta_{nn}$ (\ref{nn}) give the smaller value than $2\Delta_n^{(3)}$ for even nucleon number:  
\begin{equation} \label{sen2}
\Delta_{nn}(N)=
\begin{cases}
 G \Omega & \text{for $N=2n$,} \\
 G \Omega + G & \text{for $N=2n+1$.}
\end{cases}
\end{equation}
In this extreme case, the pairing value is $\Delta_{nn}(N) = 2\Delta_n(N+1)$ and does not explicitly depend on $N$.
\label{sec:int}
\section{Ground state multiplet}
Nucleon pairing leads to a formation of low-lying exited states with even spin values, so-called ground state multiplet (GSM). In the case of the pair of identical nucleons over a double- closed shell the degeneracy of $J \ne 0$ levels may be removed using the local $\delta$-potential  \cite{de-shalit}. 
\begin{figure}[bp!]
\includegraphics[width=65mm]{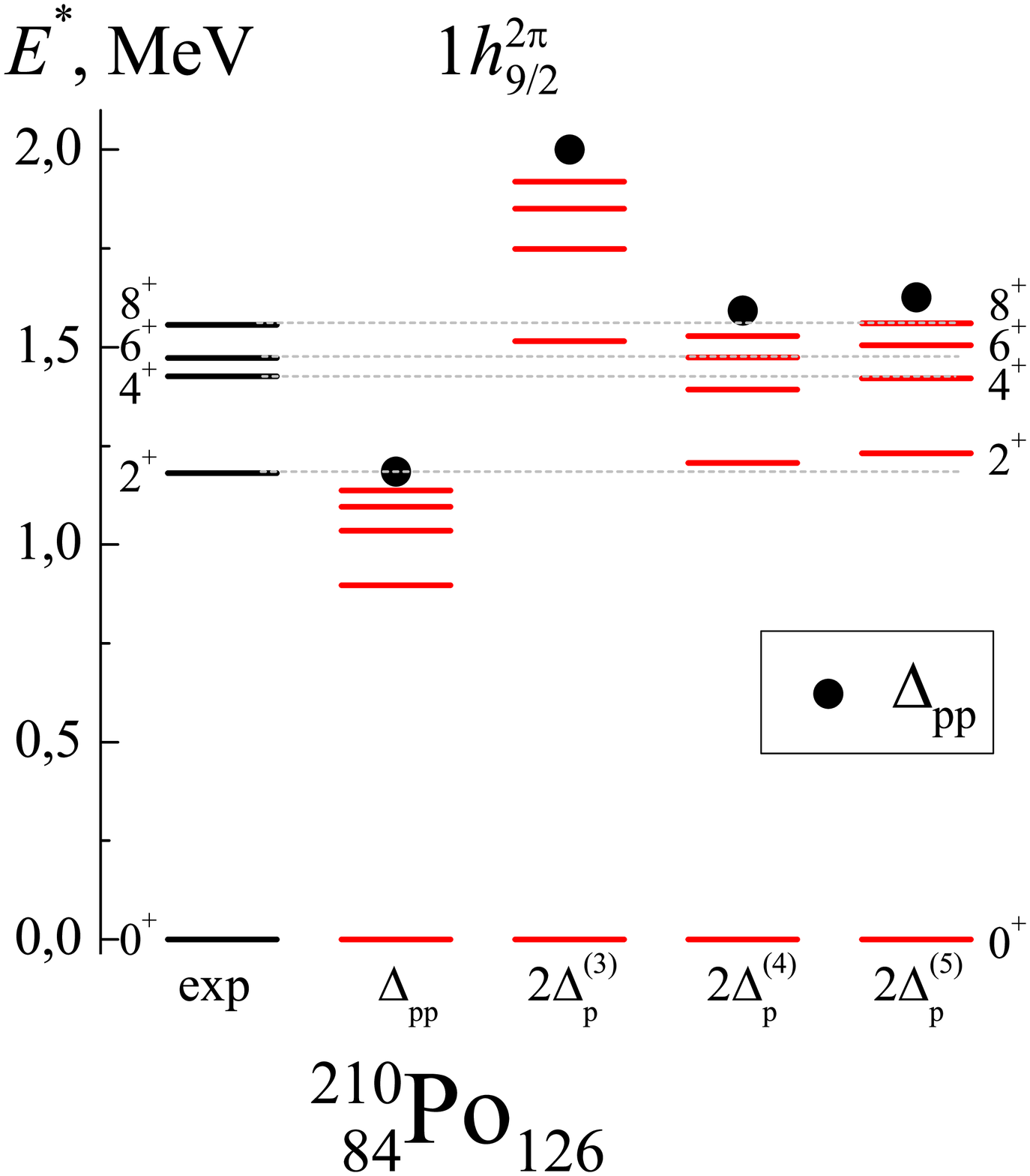}
\hfill
\includegraphics[width=65mm]{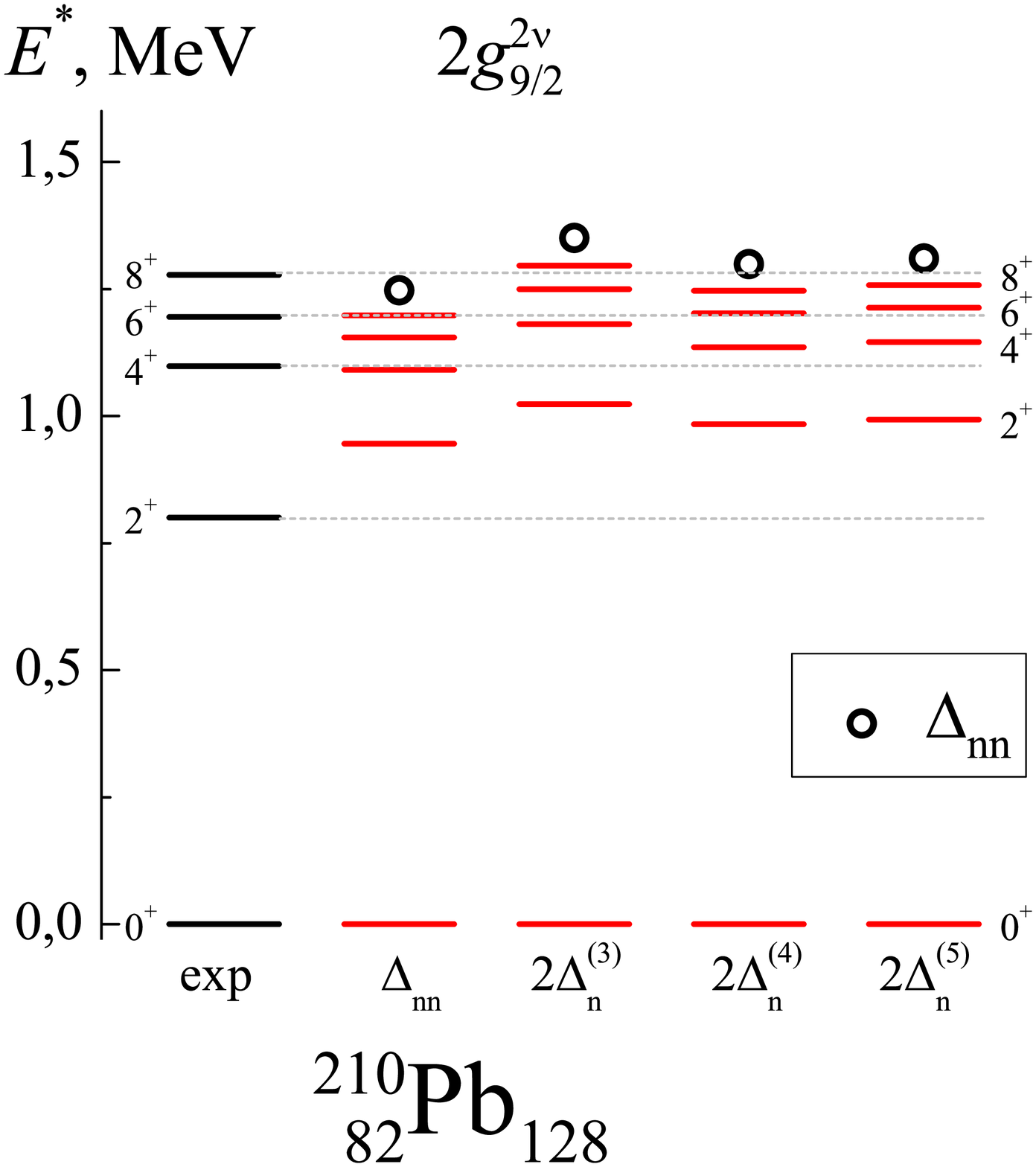}
\vspace{-3mm}
\caption{Experimental spectra \cite{exp1} and results of GSM calculation for $^{210}\rm Po$ and $^{210}\rm Pb$. Data for calculation $\Delta_n$ and $\Delta_p$ are from ~\cite{audi}.} \label{pbpo}
\vspace{-5mm}
\end{figure}
Energy levels of multiplet with $v=2$ can be found from the expression for relative energy shift: 
\begin{equation}
\frac{\Delta E_J}{\Delta E_0}=(2j+1)
\begin{pmatrix} 
j & j & J\\ 
1/2 & -1/2 & 0 
\end{pmatrix} ^2,
\end{equation}
 the ground state energy shift relative to the position of degenerate levels  $\Delta E_0$ is defined by pairing energy $\Delta_{NN}$ ~\cite{vestnik}. Strictly speaking, this approach is applicable to the nuclei with one pair of identical nucleons (holes) over a magic core only, i. e. near the magic numbers. 
However, according to the seniority model, multiplets of states with $v=2$ coincide for any number of nucleon pairs in the subshell. Indeed, as it was shown in the calculations of isotopes and isotones chains near the magic numbers 20, 50, 82, 126 \cite{Im1}, in filling of subshells with angular momentum $7/2 \le j \le 11/2$ the position of levels with $J\ge 4$ can be obtained in the $\delta$-approximation with $\Delta E_0 = \Delta_{nn}=2\Delta_n$. Thus, the correspondence between the calculated levels of GSM and experimental data can be considered as a guideline for  the nuclear pairing estimation quality.
\label{sec:res}
\section{Results}
\begin{figure}[bp!]
\includegraphics [width=120mm]{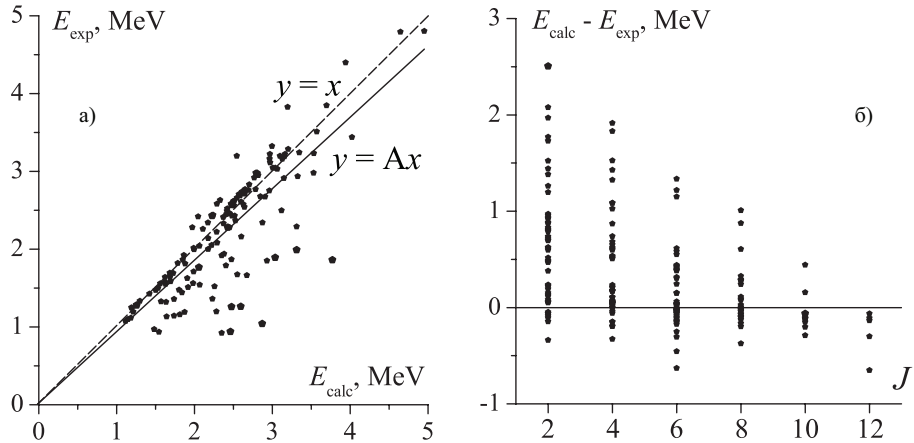}
\centering
\vspace{-3mm}
\caption{The results obtained for five-point pairing energy $\Delta E_0 = 2\Delta^{(5)}$ a) Linear approximation of compatibility between experimental and calculated energy states in GSM excluding levels with $2^+$ (right).
b) Deviation of calculated levels from experimental ones as
a function of $J$. Experimental data are from \cite{exp1}.} \label{res5point}
\vspace{-5mm}
\end{figure}
\figurename~\ref{pbpo} shows the ground state multiplets in $^{210}\rm Po$ and $^{210}\rm Pb$ for a proton (neutron) pair in the $j={9/2}$ state over the magic core $^{208}\rm Pb$. It is seen that use of pairing energy $\Delta_{nn}$ (\ref{nn}) leads to systematic underestimation of GSM levels energy, and use of EOS $2\Delta_n^{(3)}$ (\ref{three_point}) gives an overestimation of GSM levels energy. This result corresponds to the seniority model eqns. (\ref{sen2}) and (\ref{sen1}). The employment of the average values $2\Delta_n^{(4)}$ (\ref{four_point}) and $2\Delta_n^{(5)}$(\ref{five_point}) brings theory and experiment into better agreement. In this way it was considered about 50 even-even isotopes,  which main configurations can be assumed as one or several pairs of identical nucleons in a state with momentum $j \ge 7/2$ over the closed core.
\begin{table}[tbp]
\caption{The results of approximations of  pairing energy estimation methods for different sets of multiplet levels}
\begin{center}
\begin{tabular}[c]{cccccccccc}
\hline
	&  & \multicolumn{2}{c}{$J>0$} &  & \multicolumn{2}{c}{$J>2$}&  & \multicolumn{2}{c}{$J=J_{\max}$} \\
\hline 
  &  & $A$ & $\sigma$ &  & $A$ & $\sigma$ &  & $A$ & $\sigma$ \\
\hline
$\Delta_{NN}$ &  &	$0.991$&	$0.621$ &  & $1.045$ & $0.550$ &  & $1.163$ & $0.560$ \\
$2\Delta_{N}^{(3)}$ &  &	$0.751$ &	$0.902$ &  & $0.799$ & $0.749$ &  & $0.849$ & $0.608$ \\
$2\Delta_{N}^{(4)}$ &  &	$0.863$ &	$0.647$ &  & $0.914$ & $0.498$ &  & $0.991$ & $0.278$ \\
$2\Delta_{N}^{(5)}$ &  &	$0.871$ &	$0.634$ &  & $0.923$ & $0.488$ &  & $1.004$ & $0.246$ \\
\hline
\end{tabular}\label{res_table}
\end{center}
\end{table}

As the guideline of each calculation method the linear approximation of correlation the experimental and calculated energy states of GSM was used: $E_{exp} = A E_{calc}$. \figurename~\ref{res5point},~a) shows the example of  approximation for $2\Delta_n^{(5)}$. Since in medium and heavy nuclei the low-lying energy levels  with small $J$, first of all with $J^P = 2_1^+$, has  a collective interpretation, approximation without states $2^+$ was considered.
On \figurename~\ref{res5point},~b) the deviation of calculated energies from experimental ones $E_{calc} - E_{exp}$ as a function of $J$ is shown. It is clear that the deviation decreases with increasing of $J$. 
As an limit case, approximations for $J=J_{\max}$ states were considered too.

The results are shown in the table \ref{res_table}. 
In the case of full multiplet  the value of $A$ coefficient is closest to unit for $\Delta_{nn}$ with a sufficiently high value of the standard deviation $\sigma$. 
Without taking into account $2^+$ and for the case of $J=J_{\max}$ only, the doubled value of EOS effect based on masses of five adjacent nuclei $2\Delta_n^{(5)}$ gives the best estimation of multiplet splitting.


\end{document}